\documentclass{aa}   

\usepackage{natbib}
\usepackage{graphicx}
\usepackage{txfonts}
\usepackage{hyperref}
\begin{document}

   \title{Stellar halo density with LAMOST K and M giants}

   \subtitle{}
\author{M. L\'opez-Corredoira$^{1,2,3}$, X.-C. Tang$^{4}$, H. Tian$^{5,6}$,
H.-F. Wang$^7$, G. Carraro$^7$, C. Liu$^{6,8}$}
   
   \offprints{martin@lopez-corredoira.com}

\institute{$^1$ Instituto de Astrof\'\i sica de Canarias, E-38205 La Laguna, Tenerife, Spain\\
$^2$ PIFI-Visiting Scientist 2023 of Chinese Academy of Sciences at Purple Mountain Observatory, Nanjing 210023, and National Astronomical Observatories, Beijing 100012 \\
$^3$ Departamento de Astrof\'\i sica, Universidad de La Laguna, E-38206 La Laguna, Tenerife, Spain \\
$^4$ School of Physics and Astronomy, China West Normal University, 1 ShiDa Road, Nanchong 637002, China \\
$^5$ Institute for Frontiers in Astronomy and Astrophysics, Beijing Normal University, Beijing 102206, China \\
$^6$ Key Laboratory of Space Astronomy and Technology, National Astronomical Observatories, Chinese Academy of Sciences, Beijing 100101, China \\
$^7$ Dipartimento di Fisica e Astronomia, Universit\`a di Padova, Vicolo dell’Osservatorio 3, 35122 Padova, Italy\\
$^8$ University of Chinese Academy of Sciences, Beijing 100049, China 
}

   \date{}

  
  \abstract
  {}
   {We derive the morphology of the stellar component in the outer halo volume, 
and search for possible overdensities due to substructures therein.}
{We made use of some of the data releases of the spectroscopic
survey LAMOST DR8-DR9 in tandem with distance determinations for two subsamples, that is, of K-giants and M-giants, respectively, making up 60\,000 stars. These distance are obtained through Bayesian techniques that derive absolute magnitudes as a function of measured spectroscopic parameters.
 Our calculation of the density from these catalogues requires: 
(1) derivation of the selection function; 
and (2) a correction for the convolution of the distance errors, which
we carried out with Lucy's inversion of the corresponding integral equation.}
{The stellar density distribution of the outer halo (distance to the Galactic centre, $r_G$, 
of between 25 and 90 kpc) is a smooth monotonously decreasing function with a 
dependence of approximately $\rho \propto r_G^{-n}$, with $n=4.6\pm 0.4$ 
for K-giants and $n=4.5\pm 0.2$ for M-giants, and with a insignificant oblateness. 
The value of $n$ is independent of the angular distance to the Sagittarius tidal stream plane, which is what would be expected if such a
stream did not exist in the anticenter positions or  had a negligible imprint in the density distribution in the outer halo.
Apart from random fluctuations or minor anomalies in some lines of sight, we do not see substructures 
superimposed in the outer halo volume within 
the resolution that we are using and limited by the error bars. This
constrains the mass of over- and under-densities in the outer halo to be of
$\lesssim 10^3$ M$_\odot $/deg$^2$, whereas
the total mass of the stellar halo, including inner and outer parts,
is $\sim 7\times 10^8$ M$_\odot $.}
{}

\titlerunning{Halo density / LAMOST}
\authorrunning{L\'opez-Corredoira et al.}

\maketitle
%

\section{Introduction}

The stellar halo density distribution has been analysed many times
\citep[e.g.][]{You76,Fen89,Jur08,Bil08,Dea14,Xue15,Xu18,Her18,Tho18,Fuk19,Wu22}.
The representation of this component is usually given by a smooth density
function monotonously decreasing for increasing Galactocentric distances,
and with some possible oblateness.

Furthermore, some recent studies have pointed out the superposition
of some substructures \citep{Hel20}, large overdensities on the sky, and many narrow streams
\citep{Ber16,Shi18,Han22,Wu22} tentatively associated with the tidal streams of corresponding
passages of satellites. These discoveries were motivated by prior cosmological
hypotheses within the $\Lambda $CDM model, in which halos
are mostly formed by accretion and merger events, encouraging astronomers
to find structures similar to those predicted by simulations.
However, there is no information on the distance
of most of these overdensities and it is not yet clear whether these
substructures correspond to a small or negligible number of stars as fluctuations
embedded in the main field of the halo or to a major part
of the stellar component at large Galactocentric distances.

An accurate distance determination is essential for studying the morphology in the outer Galaxy,
which is not reachable with Gaia parallaxes; in any case, Gaia distances estimated with the \citet{Bai21} method
are not useful for Galactic structure analyses because this method is dependent on  
assumptions for the density distribution of the Galaxy. An interesting possibility is to employ
variable stars such as RR-Lyrae as standard candles \citep[e.g.][]{Her18}.
Distance determinations from colour and photometric metallicities \citep[e.g.][]{Hua23}
are moderately reliable, but are not as accurate as spectroscopic distances.
The available spectroscopic surveys 
and the most recent calibrations of the distance of far away sources therefore  allow us
to better constrain the morphology of the outer halo.

To this end, in this study we present an analysis that makes use of some of the 
latest data releases of the LAMOST survey \citep{Yan22},
and distance determinations of two subsamples, namely of K-giants and M-giants. We apply
Bayesian techniques in an analysis of these data to derive absolute magnitudes as a function of measured spectral parameters. 
Details of the data used in this paper are given in \S \ref{.data}.

Calculation of the density from a given catalogue is not straightforward.
There are two major technical problems to overcome in this pursuit, which are (1) the calculation of the ratio of the number of stars in our catalogue
with respect to the total number of a given type, for which we hope to derive the
selection function (see \S \ref{.sel}); 
and (2) the correction for the convolution of the distance errors (see \S \ref{.lucy}).

We present the application of our method to LAMOST K and M giants in Sects. \ref{.total} and 
\ref{.los}. In Sect. \ref{.halo}, we compare the results with theoretical models of the halo to derive the power law that best fits the data of the outer halo. In Sect. \ref{.concl}, we provide a
discussion and conclusions.

\section{Data}
\label{.data}

The LAMOST survey \citep{Yan22} covers the almost complete sky area with declination 
of between -10 and +60 deg. Here, we use subsamples derived from data releases 8 and 9, respectively, which collectively account for
around 60\,000 stars.

\subsection{M-giants}

A sample of LAMOST-DR9 M-giants is taken from \citet{Qiu23}, who use the Bayesian 
method developed by \citet{Zha20} to obtain the distance of more than 43\,000 stars from
the measured spectral parameters. After removing stars with the parameters $|Kabs_D-Kabs_M|>0.01$ 
(where $Kabs_D$ and $Kabs_M$ are the K band absolute magnitude derived from the distance and 
the model respectively; see \citet{Qiu23}), we are left with 40\,973 stars.
Figure \ref{Fig:coverage} shows the distribution of these stars in the sky.

Although referred to as M-giants, the range of spectral types in reality includes giants ($-0.4<\log g < 2.5$) between K3 and M3 spectral type, according to the selected temperatures (3200 K$< T_{\rm eff} < 4300$ K). The relative abundance of these spectral types is similar in the halo and the disc \citep{Wai92}, 
and therefore, in principle, we expect that most of these stars at large Galactocentric distance are 
genuine halo stars. 
The range of metallicities of these stars is $-1.5<[M/H]<0.5$, a range not introduced by us
but characteristic of the training samples of the SLAM (Stellar LAbel Machine) algorithm used to generate the catalogue of \citet{Qiu23} from LAMOST.
Within this constraint on metallicity, approximately half of the halo stars get removed \citep{Lop18}. Still,
half of the halo stars remain, together with disc stars and possible tidal streams. This is taken into account in the comparison with models presented  below.

\subsection{K-giants}

\begin{figure}[htb]
\vspace{1cm}
{
\par\centering \resizebox*{8.5cm}{8.5cm}{\includegraphics{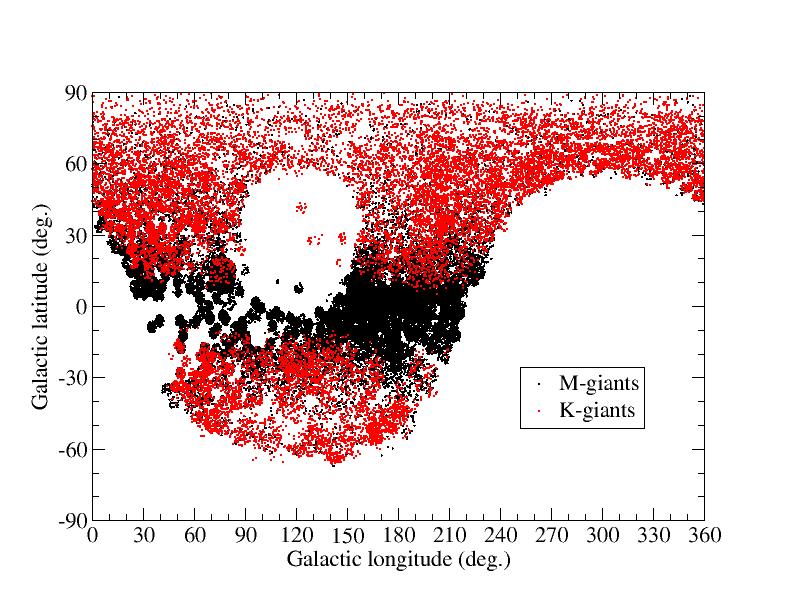}}
\par\centering
}
\caption{Sky distribution of LAMOST sources used in this paper: 
M-giants are colour-coded  in black, K-giants in red.}
\label{Fig:coverage}
\end{figure}

The sample of LAMOST-DR8 K-giants is taken from \citet{Zha23}, who use the Bayesian 
method developed by \citet{Xue14} (and applied to SDSS-SEGUE) 
to obtain the distance of 19\,521 stars from
the measured spectroscopic parameters. 
Figure \ref{Fig:coverage} shows the distribution of these stars in the sky, which is the same as for the
M-giants, because they are derived from the same LAMOST survey, but avoiding the Galactic plane.

Although indicated as K-giants,  the range of spectral types includes giants 
between G1 and K4, according to the selected temperatures (4000$<T_{\rm eff}<$5600 K).
The selection criteria are: $|z|>5.0$ kpc; [Fe/H]$<-1.0$ if 2.0 kpc$<|z|<$5.0 kpc; 
and the error in distance is lower than 30\%. 
This set of criteria filter out almost all of the stars of the disc, and yield an almost 
complete sampling of the halo stars
for $|z|>2$ kpc (once corrected for the selection function). 
We assume that the number of halo stars with [Fe/H]$\ge -1.0$ is negligible. 
The maximum heliocentric distance is over 120 kpc, which we take here as a limit.

\section{Selection function correction}
\label{.sel}

The selection function for each line of sight of Galactic coordinates $\ell$, $b$ and heliocentric distance $r$
is defined as
\begin{equation}
\label{factorS}
S(r;\ell, b)=\left\langle \frac{N_{\rm param.}(\vec{r})}{N_{\rm LAMOST}(\vec{r})}\right\rangle _{r; \ell, b} 
\left\langle \frac{N_{\rm LAMOST}(\vec{r})}{N_{\rm phot.}(\vec{r})}\right\rangle _{r; \ell, b} 
\end{equation}\[\times
\left\langle \frac{N_{\rm phot.}(\vec{r})}{N_{\rm total}(\vec{r})}
\right\rangle _{r; \ell, b} 
.\]
The first two factors \citep{Che18,Wan18} account for the 
ratio of stars with measured parameters among LAMOST sources, and the number of stars with LAMOST spectroscopy
($N_{\rm spec.}(\vec{r})$) observed at position $\vec{r}$ with respect to the total photometric sources
at 2MASS $N_{\rm phot.}(\vec{r})$,  respectively. 

For the calculation of the first two factors, 
we employ the Bayesian method as mentioned by \citet{Liu17}. We obtain the position information (right ascension (RA) and declination (Dec)) for each plate from LAMOST and use Astroquery to retrieve 2MASS photometric data within a 20 deg$^2$ region centred on that location. By combining the distribution of photometry and spectroscopy in colour--magnitude diagrams (CMDs), we derive  
$\frac{N_{\rm param.}(c,m,\ell,b)}{N_{\rm phot.}(c,m,\ell, b)}$, 
including the dependence on colour $c$ and magnitude $m$. All CMD bins have a size of $\Delta c = \Delta(J-K) = 0.1$ dex and $\Delta K = 0.25$ dex, and each CMD contains 2806 2D colour--magnitude bins. Finally, we stored the selection functions for 5533 plates of LAMOST DR9, with each plate having 2806 selection coefficients, and did the same with the plates of LAMOST DR8.

\begin{figure}[htb]
\vspace{1cm}
{
\par\centering \resizebox*{8.5cm}{8.5cm}{\includegraphics{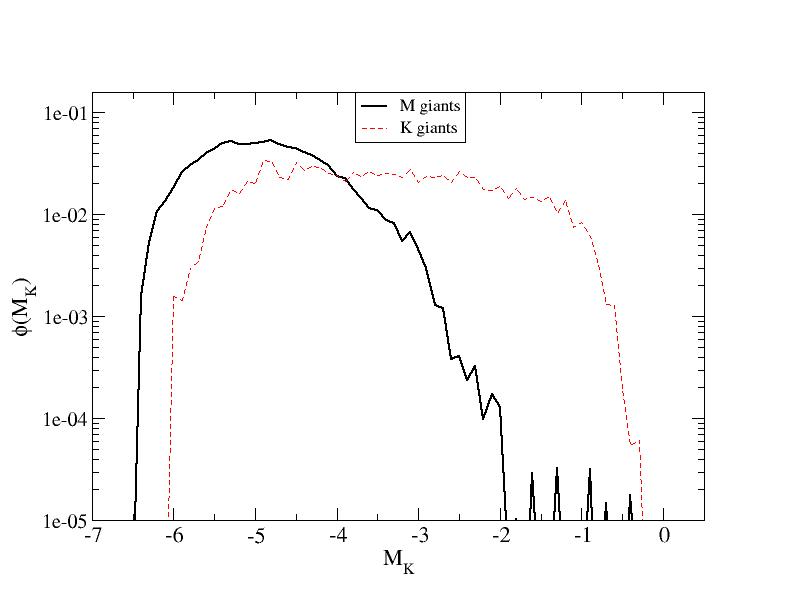}}
\par\centering
}
\caption{Luminosity function of our two samples.}
\label{Fig:lumfunc}
\end{figure}

\begin{figure}[htb]
\vspace{1cm}
{
\par\centering \resizebox*{8.5cm}{8.5cm}{\includegraphics{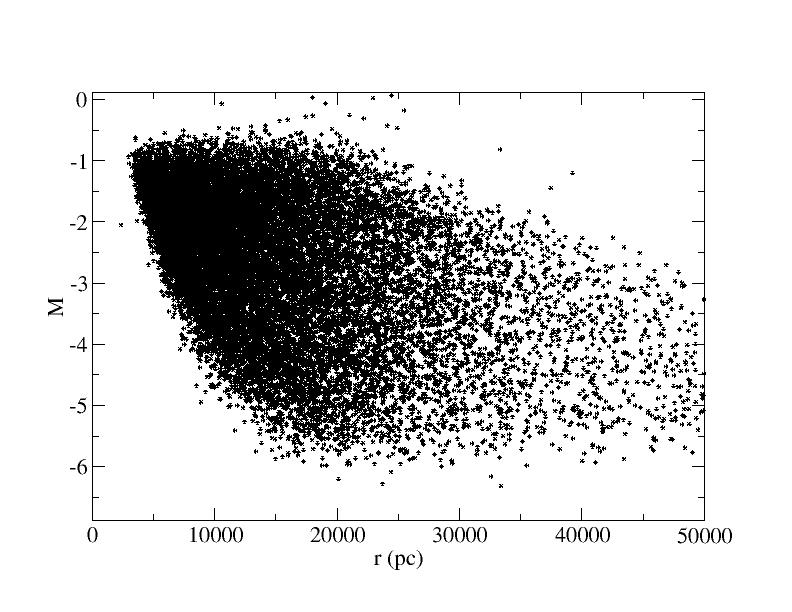}}
\vspace{1cm}
\par\centering \resizebox*{8.5cm}{8.5cm}{\includegraphics{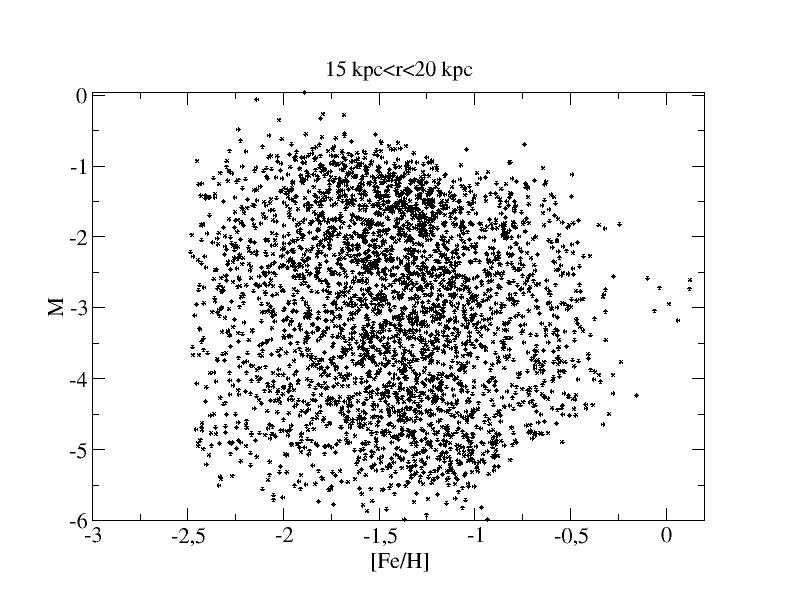}}
\par\centering
}
\caption{Absolute magnitude of the K-giant sources as a function of their heliocentric distance, and as a function
of [Fe/H] for the subsample with $r$ between 15 and 20 kpc.}
\label{Fig:Kgiants_M}
\end{figure}

The third factor in Eq. (\ref{factorS}) is an estimation of the completeness of the photometric
survey 2MASS with respect to the real distribution due to the upper magnitude limit; we calculate it as:
\begin{equation}
\left\langle \frac{N_{\rm phot.}(\vec{r})}{N_{\rm total}(\vec{r})}\right\rangle _{r; \ell, b}
=\frac{\int _{-\infty }^{M_{K,{\rm lim}}(r;\ell, b)}dM'\,\phi(M')}
{\int _{-\infty }^{\infty}dM'\,\phi(M')}
,\end{equation}\[
M_{K,{\rm lim}}(r;\ell,b)=m_{K,{\rm lim.}}-5\log_{10}[r({\rm pc})]+5-A_K(r;\ell,b)
,\]
where $m_{K,{\rm lim.}}=14.3$ is the K limiting magnitude of the 2MASS survey (with completeness 100\% and 10$\sigma $ detection), 
$A_K$ is the extinction in K-band \citep{Gre19}, and $\phi (M)$ is the luminosity function
of our stars (see Fig. \ref{Fig:lumfunc}). We note that we use the parameters of the 2MASS survey in the evaluation of this third factor
instead of the LAMOST survey, because the first two factors of completeness were calculated with respect to the 2MASS survey; however, we
detect  sources in LAMOST that are fainter than $m_{K,{\rm lim.}}=14.3$.
We derive the luminosity function from our sample within 2 kpc$<r<5$ kpc 
for M-giants, and within 15 kpc$<r<20$ kpc for K-giants and assume that we can extrapolate these to longer distances.
In these ranges of heliocentric distances, our sample covers the whole range of absolute magnitudes: 
we illustrate this in Fig. \ref{Fig:Kgiants_M} for K-giants, where we can see that, for $r<15$ kpc, there is a
lack of the brightest stars in our sample, which we tentatively believe to be due to saturation, and for $r>20$ kpc we cannot see the faintest
stars because these are beyond the completeness limit of LAMOST (not the same as the 2MASS 100\% completeness limit). 
However, we note that, even in this range, we are not 100\% complete in 2MASS, and this subsample may be complete to approximately $>90$\%; in any case, we neglect here this $\lesssim 10$\% correction in the third factor. 
However, we take into account the first two factors of the selection function.

\citet[Fig. 4]{Xue15}, based on SDSS-SEGUE data, found that the limiting magnitude of K-giants is dependent on metallicity. 
In Fig. \ref{Fig:Kgiants_M}, we show the dependence of  absolute magnitude on metallicity within the covered ranges in our LAMOST sample, 
and we do not find a remarkable dependence, except for [Fe/H]$>-1$, which might be part of some disc contamination. We therefore do not consider 
any dependence on metallicity in our analysis.

In the solar neighbourhood, the value of $S$ varies from 0.7 to $8\times 10^{-3}$ 
in the most distant regions with $r\sim 80$ kpc for the M giants sample, 
and between 0.5 and $2\times 10^{-4}$ for the  K giants sample (up to a heliocentric distance of 120 kpc).

We calculate the average density of sources within each line of sight with solid angle $\omega $ as a function of heliocentric distance $r$:
\begin{equation}
\label{counts}
\rho _{0}(r)=\frac{N_{\rm param.}(r)dr}{S(r)\omega r^2dr}
,\end{equation}
where $N_{\rm param.}(r)dr$ is the observed number of stars with spectra and measured parameters with a heliocentric distance of between $r-dr/2$ and $r+dr/2$.
This $\rho _0$ stems from a direct estimation of the density, but it is not yet corrected for the effects of the 
convolution of errors.

\section{Lucy's method on the deconvolution of the distance error}
\label{.lucy}

Due to the errors in the distances of stars, with r.m.s. $\sigma (r)$, the observed density $\rho _0$ corresponding to the distribution of stars along a given line of sight is related to the real density $\rho$ through:

\begin{equation}
\rho _{0}(r)=\int _0^\infty \rho (t)K(r,t)dt
\label{convol}
,\end{equation}
where 
\begin{equation}
K(r,t)=A(r)\,t^2\,exp\left[-\frac{(t-r)^2}{2\sigma (r)^2}\right]
,\end{equation}
which corresponds to a Gaussian distribution of errors.
$A(r)$ stands for the normalisation such that $\int _0^\infty K(r,t)dt=1$ for all $r$.

Here, we assume that the error on the distance is 21\% for the M-giants sample, as derived by \citet{Qiu23}; that is, $\sigma (r)=0.21\,r$. 
For the K-giants sample, a power-law fit of the error of the distance as a function of the distance yields an average $\sigma (r)=0.16\,r({\rm kpc})^{0.94}$ kpc.

The error distribution is of the kind of  Fredhold integral equation of the first type with kernel $K$, and can be inverted with some iterative method, such as `Lucy's method'
\citep{Lop19}:

\begin{equation}
\rho ^{[n+1]}(r)=\rho ^{[n]}(r)\frac{\int _0^\infty \frac{\rho _{0}}{\rho _{L,[n]}(s)}K(r,t)dt}
{\int _0^\infty K(r,t)dt}
,\end{equation}
\begin{equation}
\label{rhoL}
\rho _{L,[n]}(s)=\int _0^\infty \rho^{[n]}(t)K(s,t)dt
.\end{equation}

With a few iterations (determined with the algorithm of \citet{Lop19}; when $\rho _{L,[n]}(s))\approx \rho _{0}(r)$ within the error bars, and with a minimum of 3 iterations and a maximum of 15), we get the inversion of the integral equation: $\rho (r)\approx \rho ^{[n]}(r)$. The initial iteration may be set as $\rho^{[0]}(r)=\rho _{0}(r)$, but the result of the inversion is independent of the initial iteration assumption.
Also, we note that this method is model independent, as it does not assume any priors about the shape of the density.
The results of this inversion method have been compared to those from Monte Carlo simulations in previous papers presenting
applications to the deconvolution of Gaia parallaxes \citep{Lop19,Chr20}.

\section{Example application to entire samples}
\label{.total}

\begin{figure}[htb]
\vspace{1cm}
{
\par\centering \resizebox*{8.5cm}{8.5cm}{\includegraphics{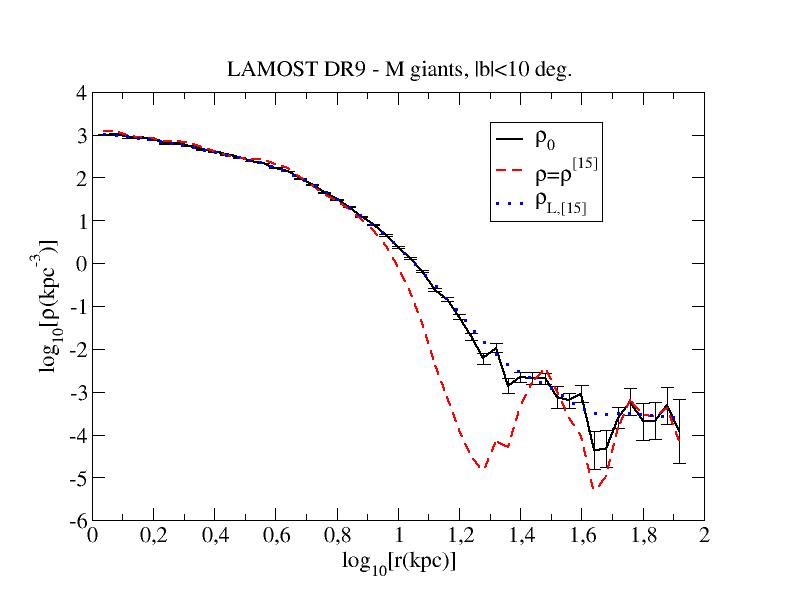}}
\vspace{1cm}
\par\centering \resizebox*{8.5cm}{8.5cm}{\includegraphics{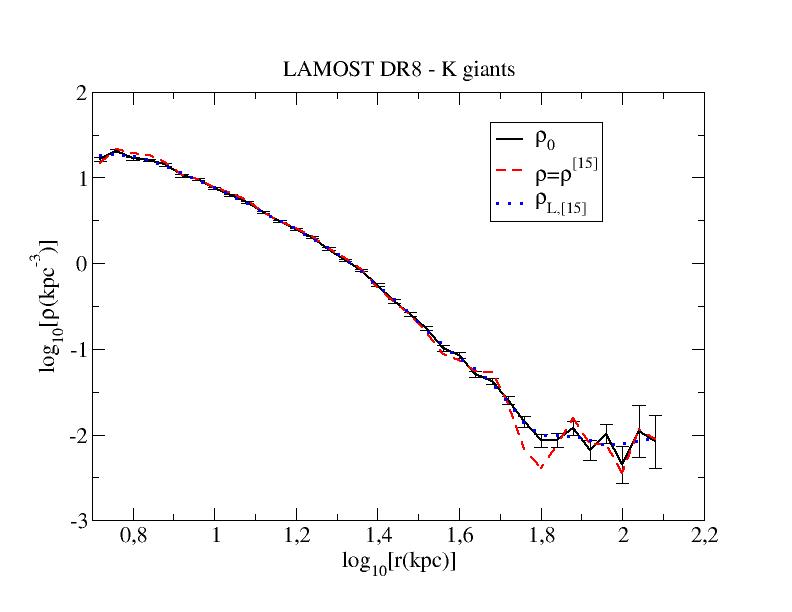}}
\par\centering
}
\caption{
Stellar density as a function of heliocentric distance for M-giants (Top panel; 40\,973 sources) and K-giants (Bottom panel; 19\,521 sources). $\rho_0$ is the observed density without correction of deconvolution or parallax errors; $\rho $ is the density with deconvolution; $\rho_{L,[15]}$ is the amount derived in Eq. (\ref{rhoL}), which converges to $\rho _0$ after 15 iterations.}
\label{Fig:density}
\end{figure}

\begin{figure}[htb]
\vspace{1cm}
{
\par\centering \resizebox*{8.5cm}{8.5cm}{\includegraphics{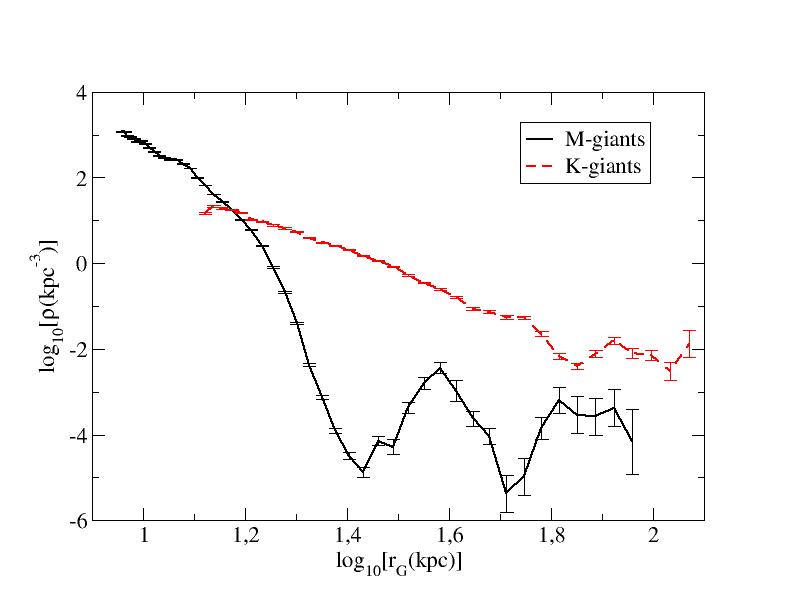}}
\par\centering
}
\caption{Stellar density after deconvolution correction as a function of the distance to Galactic centre for M-giants (Top panel; 40\,973 sources) and K-giants (Bottom panel; 19\,521 sources).
Error bars correspond to the Poissonian errors of the measured $\rho_0$.}
\label{Fig:densityG}
\end{figure}

Let us consider the average of the whole sky coverage of both samples: for M-giants,  
$\omega=6.6$ stereo-radians, assuming that all the sources with declination between 
-10 and +60 deg. are observed; for K-giants, we add the extra constraint of avoiding
regions within $|z|<2$ kpc, which gives a $\omega$ of between 4.3 and 6.5 stereo-radians, depending on the
distance. We calculate the average density of observed 
 sources as a function of heliocentric distance $r$.
The result is plotted in log--log (step: $\Delta log_{10}(r)=0.04$) in Fig. \ref{Fig:density} 
as a function of heliocentric distance and Fig. \ref{Fig:densityG} as a function of the 
distance to the Galactic centre, $r_G=\sqrt{R^2+z^2}$.

In the profile $\rho _{0}(r)$ for M-giants, we observe two ranges, that is, closer than and beyond 20 kpc ($\log_{10}r({\rm kpc})=1.3$), which might be interpreted as the volumes where the disc and halo are predominant, respectively. It is also observed that the density is monotonously decreasing.
Nonetheless, this observed density is not correct; rather it corresponds to the convolution of the real density, as expressed
in Eq. (\ref{convol}). 
When we apply the previous method of deconvolution to this $\rho _{0}(r)$,
we obtain $\rho (r)$ as shown in the red-dashed line in the upper
panel of Fig. \ref{Fig:density}. 
As expected, the density $\rho (r)$ in the outer region ($r>10$ kpc) is much lower than  $\rho _{0}(r)$.
Remarkably, one can see that the real density is not monotonously decreasing, but there are two minima of the density around $r=18$ kpc ($\log_{10}r({\rm kpc})=1.25$) and $r=45$ kpc ($\log_{10}r({\rm kpc})=1.65$), and then two substructures appear with peaks around $r=28$ and $r=63$ kpc ($\log_{10}r({\rm kpc})=1.45$, $\log_{10}r({\rm kpc})=1.80$). This shows the power of this method in recovering information on the density, and highlights some structures that were masked by the convolution of errors. 
However, these types of under-densities and over-densities only represent an average within areas in the first three quadrants;
the combination of areas with different depth might produce this kind of artefact. In order to check for the possible existence
of substructures, in the following section we examine the analysis with more accurate space resolution, separating different lines
 of sight only in the second and third quadrant, which allows us to distinguish structures not only as a function of distance but also of the  position in the sky.

For the K-giants, the effect of Lucy's deconvolution is less significant, because the error on the
distance $\sigma (r)$ is much smaller than for M-giants. This survey has removed the Galactic plane
and has a predominant contribution to the density at high latitudes. Again, there can be no direct interpretation of the under-densities
and peaks in $\rho (r)$ because we are combining many different lines of sight and we need to separate these to derive the mean density along them. Nonetheless, the exercise in this section serves to illustrate the application of the methodology.

\section{Different lines of sight}
\label{.los}

\begin{figure}[htb]
\vspace{1cm}
{
\par\centering \resizebox*{8cm}{8cm}{\includegraphics{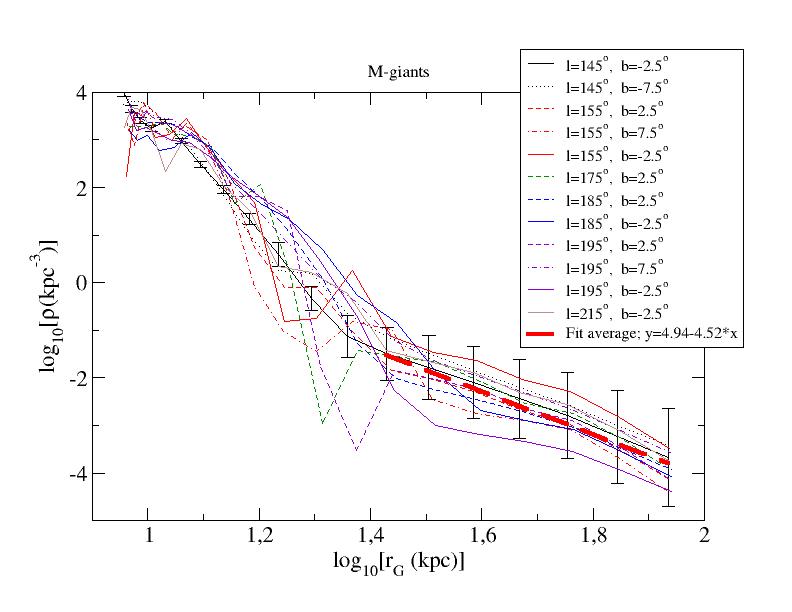}}
\vspace{1cm}
\par\centering \resizebox*{8cm}{8cm}{\includegraphics{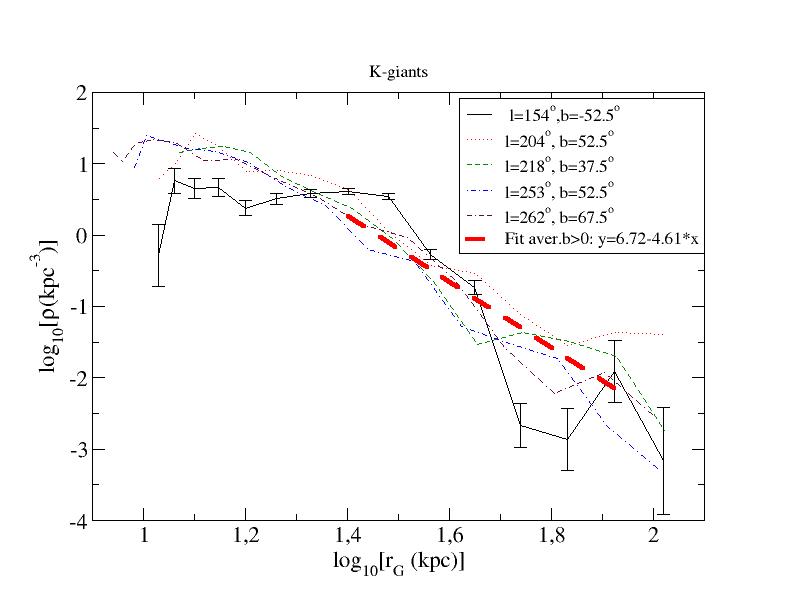}}
\par\centering
}
\caption{Stellar density after deconvolution correction as a function of the distance to
the Galactic centre.
Upper  panel: M-giants within 12 lines
of sight with $\Delta b=5^\circ $, $\Delta \ell=10^\circ$.
Bottom panel: K-giants within five lines
of sight with $\Delta b=15^\circ $, $\Delta \ell=30^\circ/cos(b)$. Fits correspond to the average
of all lines of sight in the range $\log_{10}[r_G({\rm
kpc})]$ between 1.4 and 1.95, except the line $\ell =154^\circ $, $b=-52.5^\circ $, which presents
an anomaly. Error bars (corresponding to the error of measured $\rho_0$) of the first line of sight are plotted; for other lines of sight, the error
bars are similar.}
\label{Fig:density2}
\end{figure}

We apply this method of deconvolution (together with the selection function analysis) to different lines of sight corresponding to different subsamples of LAMOST to derive the star density.
The different results of the deconvolution are shown in Fig. \ref{Fig:density2}.

For the M-giants sample,
we divide the sky within $90^\circ <\ell <270^\circ $ in regions with $\Delta b=5^\circ $, $\Delta \ell=10^\circ$; the area of these regions is 50$\cos (\overline{b})$ square degrees when they are totally covered by the LAMOST survey or lower otherwise. 
We consider only the areas with more than 400 stars. This results in a total of 12 regions, all of them within $|b|\le 10^\circ $, and with a covered area of larger than 47.9 deg$^2$ each. We run the Lucy's deconvolution method on them using a step of $\Delta log_{10}(r)=0.1$.

They show a monotonously
decreasing density, typical of a halo density distribution, with no significant dependence on
the latitude.  This indicates a low oblateness of the halo; the exception is the line of sight toward
$\ell =155^\circ$, $b=-2.5^\circ$, which presents a relative peak at $r_G\approx 23$ kpc.

For the K-giants sample, we have a lower number of stars per square degree, and so we take larger areas:
we divide the sky within $90^\circ <\ell <270^\circ $ in regions with $\Delta b=15^\circ $, $\Delta \ell=30^\circ /cos(b)$. We only consider the areas with more than 400 stars: a total of five regions, all of them off-plane, with a covered area of each line of sight of between 170 and 360 deg$^2$. We run the Lucy's deconvolution method on them using a step of $\Delta log_{10}(r)=0.1$.
 In general, the density of these areas shows a similar dependence on the distance to that of the centre of the Galaxy: a monotonously
decreasing density typical of a halo density distribution, with no significant dependence on
the latitude, which again indicates a low oblateness of the halo; the exception is the line of sight toward
$\ell =154^\circ$, $b=-52.5^\circ$, which exhibits a relative peak at $r_G\approx 25$ kpc.

For K-giants, the five selected lines of sight are clearly separated and do not allow us to build a
3D map. For the M-giant lines of sight, which are more continuous and confined in the plane,
we combine the derived densities of the 12 of them to produce a 3D map in Galactocentric coordinates
(we assume $R_\odot =8$ kpc and  we neglect the height of the Sun over the plane: $Z_\odot =0$).
In Fig. \ref{Fig:slices}, we plot three slices of this 3D map parallel to the XY plane, 
corresponding to $-6$ kpc$<Z<-2$ kpc, $|Z|<2$ kpc, and
$2$ kpc$<Z<6$ kpc, respectively. In Fig. \ref{Fig:2Ddenozl5} we plot the density $\rho _0$ (without deconvolution)
only for the plane region  for
comparison; and in Fig. \ref{Fig:counts} we show the corresponding counts of stars per unit volume.

These maps indicate a smooth distribution without clear substructures above the noise level.
There are no significant over-densities and the only remarkable under-density is for 
$Y=-20$ to -10, $X=40$ to 80, $Z=-6$ to 0 (units are in kpc), which corresponds to just one line
of sight $\ell =195 ^\circ $, $b=-2.5^\circ $. Figure \ref{Fig:density2} shows how the
density along this line of sight for all distances is much lower than the density in the
other lines of sight, but with the same monotonously decreasing power-law shape. Given
that there are no signs of structure with peaks or valleys along this line of sight, we attribute
the global under-density for this line of sight to some possible loss of stars or miscalibration
of extinction or the selection function in LAMOST.

The maps of Fig. \ref{Fig:slices} show that the density 
of stars after applying Lucy's deconvolution 
at the largest $R$ is much lower than that without this correction:
from Fig. \ref{Fig:density2} (M-giants
at $R>40$ kpc), this density is $\lesssim 2.5\times 10^{-3}$ kpc$^{-3}$ ($\lesssim 10^{-6}$ times the solar neighbourhood density);
at $R>70$ kpc, it is $\lesssim 5\times 10^{-4}$ kpc$^{-3}$ ($\lesssim 
2\times 10^{-7}$ times the solar neighbourhood density).
Two orders of magnitude larger densities are found for K-giants in off-plane regions 
for similar Galactocentric distances ($r_G=\sqrt{R^2+z^2}$), which we attribute to the larger
number of K-giants than M-giants.

The observed density of M-giants in the solar neighbourhood in Fig. \ref{Fig:density2} 
is $\rho _{\odot ,M-giants} \sim 2.5\times 10^3$ star/kpc$^3$ (this is dominated by disc stars). 
For K-giants, we have no direct density
measurements in the solar neighbourhood (the Galactic plane stars were removed), but we
estimate a ratio of K-giants/M-giants of $\sim 10^{1.6}$ (derived as the average
from Fig. \ref{Fig:density2} in the range of $\log r_G=1.4-1.9$, where the disc does 
not contribute and neglecting the oblateness of the halo), and so this would imply $\rho _{\odot ,K-giants} \sim 10^5$ star/kpc$^3$.
 
\section{Halo density}
\label{.halo}

Most of the regions with $R>25$ kpc or $|z|>4$ kpc should be explained in terms of halo density. For comparison,
Fig. \ref{Fig:halomodel} shows the prediction of the density of stars in the  halo following the model by \citet{Fen89} and \citet{Bil08}:
\begin{equation}
\rho _{\rm halo}=1.4\times 10^{-3}f\,\rho_\odot
 \frac{\exp[Q (1-X_{sp}^{0.25})]}{X_{sp}^{0.875}}
,\end{equation}\[
X_{sp}=\frac{\sqrt{X^2+Y^2+(Z/q)^2}}{R_\odot }
,\]
where $Q=10.093$ and $q=0.63$. In the case of M-giants, 
we add an extra factor $f=0.5$ in the normalisation to take into account the fact that in the observed range of metallicities we
include all the M-giants of the disc but only $\sim 50$\% of the M-giants in the halo \citep{Lop18};
for the K-giants sample, we take $f=1$.

For M-giants, comparing Fig. \ref{Fig:halomodel} with the central
panel of Fig. \ref{Fig:slices}, we see that \citet{Fen89} and \citet{Bil08} predict a similar density to that observed 
at large Galactocentric distances. We can therefore say that most of the stars at $X>25$ kpc can be
explained in terms of the stellar halo and there is no need for extragalactic components or multiple substructures. 
Of course, at $X<25,$ the halo model gives much lower stellar
density than observed because the disc dominates in that volume.
 
Another model derived by \citet{Xu18} provides a parametrisation of
\begin{equation}
\label{Xu18}
\rho _{\rm halo}=A_{X,\odot }\, X_{sp}^{-n}
,\end{equation}
where $n=5.03$ and $q\approx 1$ for $X_{sp}>4$. With this new profile, and adopting the same normalisation as above,
$A_{X,\odot }=1.4\times 10^{-3}f\,\rho_\odot $, we get the density distribution plotted in Fig. \ref{Fig:halomodel2}.
This second model yields much lower densities of M-giants 
than our data in the outer parts.
Instead, with this simple law of Eq. (\ref{Xu18}), we would need $n=4.52\pm 0.21$ with M-giants and 
$n=4.61\pm 0.36$ with K-giants (see fits of Fig. \ref{Fig:density2} in the range $\log_{10}[r_G({\rm
kpc})]$ between 1.40 and 1.95). This value of $n$ is very similar to the one obtained in other previous analyses
\citep[e.g.][]{Xue15,Her18}, although, as opposed to these latter authors, 
we do not find clear fluctuations or variations indicative of possible substructures.
In Table \ref{Tab:n}, we give the values of $n$ for each line of sight, and
in Fig. \ref{Fig:residuals}, we show the residuals with respect to these fits in that range of $r_G$. In the case
of M-giants, these are within Poissonian errors; in the K-giants, fluctuations are larger than Poissonian errors but still
small and random, without any clear defined structure.
In Fig. \ref{Fig:slope_sky}, we show a plot that is similar to the bottom panel of Fig. 13 of \citet{Her18}, with variation of the average 
slope $n$ with the sky position in the 17 investigated lines of sight. This variation of $n$ appears to be random and not
associated to particular regions of the sky. In general, apart from the anomalies already pointed out in the previous section, 
no significant coherent variations of the slope or fluctuations are observed.

Of these 17 lines of sight, 7 have their centre at an angular distance of lower than 10 degrees from the main Sagittarius stream orbit,
which might affect the determination of the slope of the smooth halo profile \citep{Tho18}. If we calculate the power-law index $n$ only with
the lines of sight with $|\widetilde{B}|>10$ deg., we get $n=4.50\pm 0.18$ for M-giants and $n=4.57\pm 0.56$ for K-giants. No significant difference
with respect to the values of $n$ including the lines of sight close to Sagittarius stream orbit.
In Fig. \ref{Fig:Sagittarius}, we plot $n$ versus $\widetilde{B}$. We do not find any  correlation. Sagittarius stream regions provide a negligible contribution
to the change of the halo profile. We could even conclude  that our data are compatible with no detection at all of such a putative tidal stream.

\begin{table}
\caption{Best linear fits of the density profile in log--log, $log_{10}(\rho )=A-n(x-1.40)$, $x=\log_{10}[r_G({\rm kpc})]$, 
in the range $\log_{10}[r_G({\rm kpc})]$ between 1.40 and 1.95 for the different lines of sight. $(\ell, b)$ are Galactic coordinates of the central position of the line of sight; $(\widetilde{\Lambda}, \widetilde{B})$ are the Sagittarius stream coordinates \citep[appendix A]{Bel14}.}
\begin{center}
\begin{tabular}{ccccc}
Sample & $(\ell, b)$ [deg.] & $(\widetilde{\Lambda}, \widetilde{B})$ [deg.] & $A$ & $n$ \\ \hline
M-giants & (145,-2.5) & (208,37) & $-1.32\pm 0.03$ & $4.30\pm 0.09$ \\
M-giants & (145,-7.5) & (213,35) & $-1.00\pm 0.02$ & $4.48\pm 0.06$ \\
M-giants & (155,2.5) & (199,29) & $-1.53\pm 0.10$ & $4.49\pm 0.31$ \\
M-giants & (155,7.5) & (193,30) & $-1.35\pm 0.29$ & $5.53\pm 0.89$ \\
M-giants & (155,-2.5) & (204,27) & $-0.88\pm 0.11$ & $4.48\pm 0.34$ \\
M-giants & (175,2.5) & (194,9) & $-1.17\pm 0.10$ & $4.80\pm 0.31$ \\
M-giants & (185,2.5) & (191,-1) & $-1.77\pm 0.07$ & $3.79\pm 0.21$ \\
M-giants & (185,-2.5) & (196,-2) & $-1.08\pm 0.26$ & $5.72\pm 0.78$ \\
M-giants & (195,2.5) & (189,-10) & $-1.47\pm 0.13$ & $4.53\pm 0.40$ \\
M-giants & (195,7.5) & (184,-9) & $-1.06\pm 0.03$ & $4.57\pm 0.09$ \\
M-giants & (195,-2.5) & (194,-12) & $-2.33\pm 0.13$ & $3.70\pm 0.39$ \\
M-giants & (215,-2.5) & (189,-31) & $-1.15\pm 0.11$ & $4.46\pm 0.33$ \\ \hline
K-giants & (154,-52.5) & (252,6) & $0.70\pm 0.49$ & $6.82\pm 1.58$ \\
K-giants & (204,52.5) & (139,-1) & $0.42\pm 0.16$ & $4.02\pm 0.51$ \\
K-giants & (218,37.5) & (147,-17) & $0.14\pm 0.25$ & $4.01\pm 0.78$ \\
K-giants & (253,52.5) & (118,-22) & $0.10\pm 0.19$ & $5.12\pm 0.59$ \\
K-giants & (262,67.5) & (109,-9) & $0.36\pm 0.28$ & $5.32\pm 0.89$ \\ \hline
\end{tabular}
\end{center}
\label{Tab:n}
\end{table}

By performing the inverse calculation of star counts with this power-law distribution, using Eq. (\ref{counts}),
the number of observed (with spectra) stars with Galactocentric distance ($r_G$) larger than $r_{\rm G,min}$ should
be 
\begin{equation}
N(r_G>r_{\rm G,min})=\omega _G\,A_{X,\odot }\int _{r_{\rm G,min}}^\infty dr_G\,r_G^2\,(r_G/R_\odot)^{-n}S(r_G)
.\end{equation}

Assuming a selection function $S(r_G)=S(r_{\rm G,min})\left(\frac{r_G}{r_{\rm G,min}}\right)^{-\beta}$, 
$\omega_G,$ the angular area of the
sky observed from a Galactocentric position (which is not the same thing as $\omega $,  which is the area observed in heliocentric coordinates,
but they are similar at large $R$, and so we can assume $\omega _G=\omega =6.5$ stereo-radians) is equal to
\begin{equation}
\label{halostars}
N(r_G>r_{\rm G,min})\sim \frac{S(r_{\rm G,min})
\,\omega \,A_{X,\odot }R_\odot ^{n}r_{\rm G,min}^{3-n}}{\beta+n-3}
,\end{equation}
where $n=4.6$ for both samples. For our samples, we measure:
$A_{X,\odot }=1.7$ kpc$^{-3}$, S(50 kpc)=0.5, $\beta =7.6$ for M-giants; 
$A_{X,\odot }=140$ kpc$^{-3}$, S(50 kpc)=0.07,  $\beta =4.0$ for K-giants. 
With these parameters, the number of
halo stars with $r_G$ larger than 50 kpc would be $\sim 16$ M-giants and $\sim 310$ K-giants. 
These numbers are close to the observed number of stars with $r_G>50$ kpc: 10 M-giants;
273 K-giants \citep{Zha23}.

The total number of stars of the halo for $r_{\rm G,min}>4$ kpc (neglecting the oblateness terms and
including those that are not M-giants or K-giants, within $M_G<10$) 
can also be derived from Eq. (\ref{halostars}) by setting $r_{\rm G,min}=4$ kpc, 
$S(4\ {\rm kpc})=1$, $\beta =0$, 
$\omega =4\pi$, $A_{X,\odot }=1.4\times 10^{-3}\rho _{\odot }$ and assuming a total stellar 
density (including disc and halo) within $M_G<10$ of $\rho _\odot =6.4\times 10^7$ kpc$^{-3}$ \citep{Chr20}.
This results $N\sim 10^9$ stars. The stellar mass density of  solar neighbourhood is $4.3\times 10^7$ M$_\odot$ kpc$^{-3}$ \citep{McK15}, and therefore the ratio of mass per star (for stars with $M_G<10$) in the solar neighbourhood is 2/3 M$_\odot$/star.hi
This ratio is subject to some uncertainties, but the order of magnitude is not expected to change significantly.
Assuming a similar mass per star ratio in the disc and in the halo, we would have a total mass of the stellar halo at $R>4$ kpc of $\sim 7\times 10^8$ M$_\odot $. This number is similar to other values estimated in the literature, which give $4-7\times 10^8$ M$_\odot $ (review at \citet{Bla16}).
 
The stellar mass estimated with the same procedure for the outer halo ($r_G>25$ kpc) gives $\sim 4\times 10^7$ M$_\odot $. The fluctuations of the density
in the top panel of Fig. \ref{Fig:density2} are $\Delta \log_{10}\rho \sim 0.6$ within
50 square degrees (1/800 of sky area), 
which means that possible substructures in the outer halo 
within the level of the fluctuations should
have a mass of $\lesssim 6\times 10^4$ M$_\odot $, that is,
$\lesssim 10^3$ M$_\odot $/deg$^2$.

\section{Discussion and conclusions}
\label{.concl}

The outer halo stellar density distribution is a smooth monotonously decreasing function of
the distance to the Galactic centre, $r_G$, with a dependence of approximately
$\rho \propto r_G^{-n}$, with $n\approx 4.6$ for K-giants and $n\approx 4.5$ for M-giants, 
which are compatible with each another.

We did not investigate the halo oblateness \citep{Jur08} or prolateness \citep{Tho18,Fuk19} found by other authors. 
The halo should be almost spherical at the large Galactocentric distances explored in the present study \citep{Xu18}, and indeed we do not find any significant trend in the dependence of the density on Galactic latitude. Nor have we investigated asymmetries of stellar density in the halo, 
as found by other teams \citep{Xu06}.

Analyses of metallicity and kinematics reveal differences between the inner and outer halo \citep{Car07}.
There are also gradients of metallicity
in the halo stars with respect to the distance to the Galactic plane \citep{Ron01,Ak07}.
For the density analyses, we do not consider necessary to distinguish between the two halos, and we
do not separate the different populations with the different metallicities, although for the M-giants
subsample we only select the most metal-rich ones ([M/H]$>$-1.5).

While the convolution with the error function may erase some substructure, the deconvolution
produces the opposite effect: 
Lucy's method of deconvolution would recover over-densities of sufficient amplitude and size,
as shown in Sect. \ref{.total} (see also references cited in Sect. \ref{.lucy}), but we simply do not see them. 
We do not see substructures superimposed on the halo volume within the resolution used here and limited
by the error bars. As shown in Fig. \ref{Fig:density2}, we do not see over-densities beyond possible random fluctuations: only a possible exception in the 
over-density at $r_G\approx 22-25$ kpc, for K-giants and M-giants at $\ell=154^\circ $, $b<0$, but
we cannot exclude that this is a random fluctuation, and in any case it is in the volume dominated by the disc. In the halo-dominated volume 
($r_G>25$ kpc, i.e., $\log_{10}[R(kpc)]>1.4$), the density functions are quite smooth within the error bars. The distribution  of
M-giants in our maps 
does not match the expectation that it is dominated by the Sagittarius Stream, and is in disagreement with the claims by \citet{Qiu23} based on the same
sample. Also, other works with other surveys claim to have detected the Sagittarius Stream \citep[e.g.][]{Her18,Sta19}. We have not found it.
As a matter of fact, we see that the value of $n$ is independent of the angular distance to the Sagittarius tidal stream plane, which would be expected if such a
stream did not exist in the anticentre positions or had a negligible imprint on the density distribution (in the outer halo, $r_G>25$ kpc; though it may be present in the inner
halo).

We note that with our LAMOST data the number of stars is not high (20\,000 or 40\,000 for each subsample in the whole sky), and so LAMOST cannot detect the same substructures observed with SDSS or Gaia or similar surveys 
with many millions of sources.
Possible substructures in the outer halo
within the level of the fluctuations with this LAMOST survey should
have a mass of $\lesssim 10^3$ M$_\odot $/deg$^2$.

Moreover, we did not use kinematics here \citep{Hel20,Wu22}, 
nor chemical  information \citep{Hel20,Wu22,Hor23}. 
Rather, we explored the density profile, which is more direct evidence of overdensities
and  is model independent, as opposed to  kinematics- and chemistry-based selection.
We have the advantage of access to distance information, whereas most of the analyses finding over-densities in the sky only considered the projection of the substructures, 
but did not possess distance information because of their photometric errors, or because Gaia parallaxes do not reach those distances.
Some breaks in the density profile were previously identified at $r_G\approx 25$ kpc 
(e.g. Gaia-Sausage-Enceladus; \citet{Han22}), but  whether or not the inner part within $r_G<25$ kpc is  contaminated by the disc is not clear (albeit in principle excluded using metallicities).

In conclusion, from
our analysis of LAMOST sources, we see that a smooth halo may explain the observed distribution with no significant over-densities (except perhaps in 1 or 2 lines of sight among the 17 we have explored); this does not mean that there are no substructures, 
but we cannot see them with the resolution of our bins and beyond the Poissonian noise of star counts.
\\

{\bf Acknowledgements:}
Thanks are given to the anonymous referee for very useful comments that helped to improve this paper.
Thanks are given to the language editor of A\&A Johua Neve for proof-reading the text.
MLC’s research is supported by the Chinese Academy of Sciences 
President’s International Fellowship Initiative grant number 2023VMB0001
and the grant PGC-2018-102249-B-100 of the Spanish Ministry of Economy and Competitiveness (MINECO).
HT is supported by Beijing Natural Science Foundation with
Grant No. 1214028 and the National Natural Science Foundation of China (NSFC) under grant 12103062.
HFW is supported in this work by the Department of Physics and Astronomy 
of Padova University though the 2022 ARPE grant: {\it Rediscovering our Galaxy with machines.}
CL thanks the NSFC with
grant No.11835057 and the National Key R\&D Program of
China No. 2019YFA0405501.
Guoshoujing Telescope (the Large Sky Area Multi-Object Fiber
Spectroscopic Telescope LAMOST) is a National Major Scientific Project built by the Chinese Academy of
Sciences. Funding for the project has been provided by the National Development and Reform Commission.
LAMOST is operated and managed by the National Astronomical Observatories, Chinese Academy of
Sciences.

\section*{ORCID for authors}

M. L\'opez-Corredoira: 0000-0001-6128-6274

X. C. Tang:  0009-0004-5137-0092

H. Tian: 0000-0003-3347-7596

H.-F. Wang: 0000-0001-8459-1036

G. Carraro: 0000-0002-0155-9434

C. Liu: 0000-0002-1802-6917

\appendix

\section{Other figures of Sects. \ref{.los}, \ref{.halo}}

\begin{figure}[htb]
\vspace{1cm}
{
\par\centering \resizebox*{6.5cm}{6.5cm}{\includegraphics{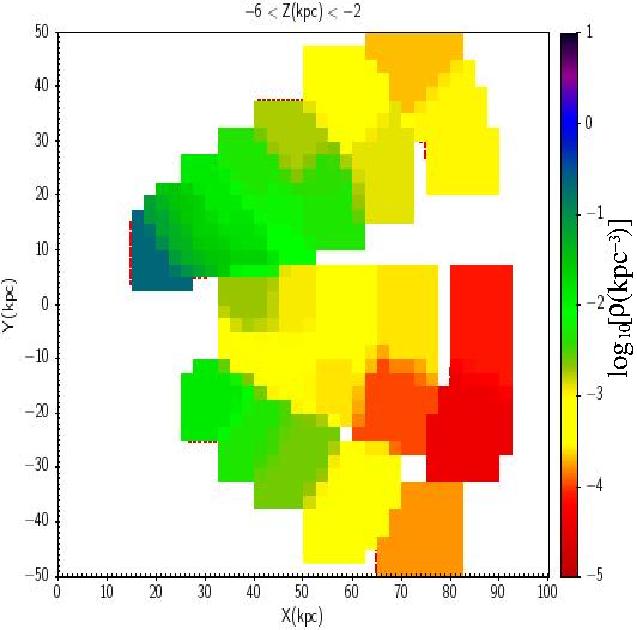}}\\
\vspace{1cm}
\par\centering \resizebox*{6.5cm}{6.5cm}{\includegraphics{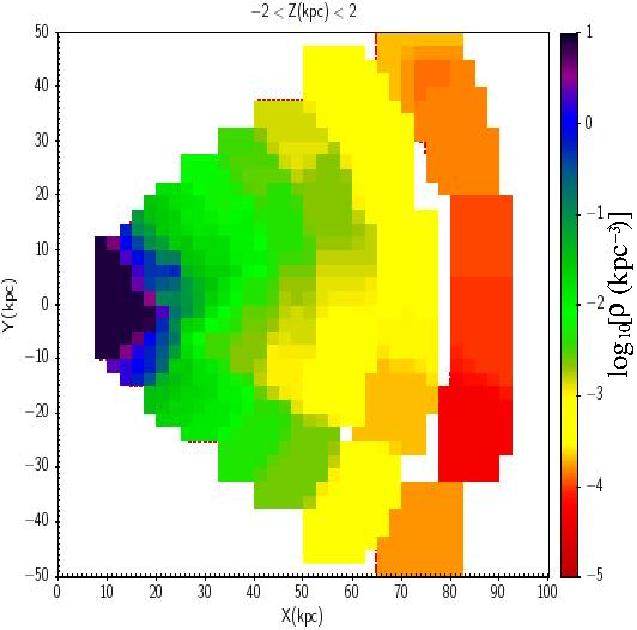}}\\
\vspace{1cm}
\par\centering\ \resizebox*{6.5cm}{6.5cm}{\includegraphics{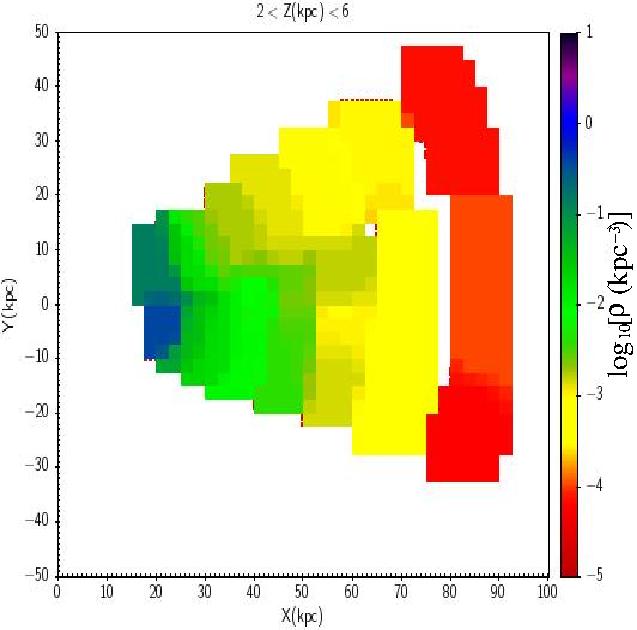}}
\par\centering
}
\caption{Density of M-giants with deconvolution of distance errors. Pixel size 1x1 (kpc); interpolation in X and Y directions up to 5 pixels.}
\label{Fig:slices}
\end{figure}

\begin{figure}[htb]
\vspace{1cm}
{
\par\centering \resizebox*{8.5cm}{8.5cm}{\includegraphics{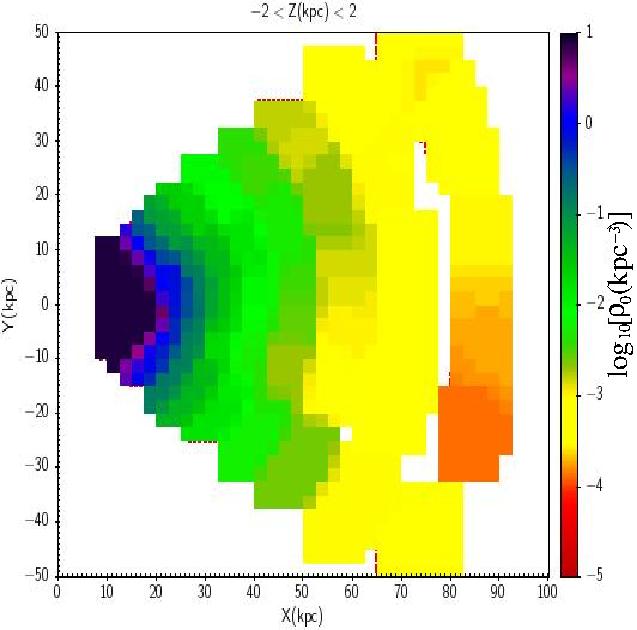}}
\par\centering
}
\caption{Density of M-giants without deconvolution correction ($\rho _0$). 
Pixel size 1x1 (kpc); interpolation in X and Y directions up to 5 pixels.}
\label{Fig:2Ddenozl5}
\end{figure}

\begin{figure}[htb]
\vspace{1cm}
{
\par\centering \resizebox*{8.5cm}{8.5cm}{\includegraphics{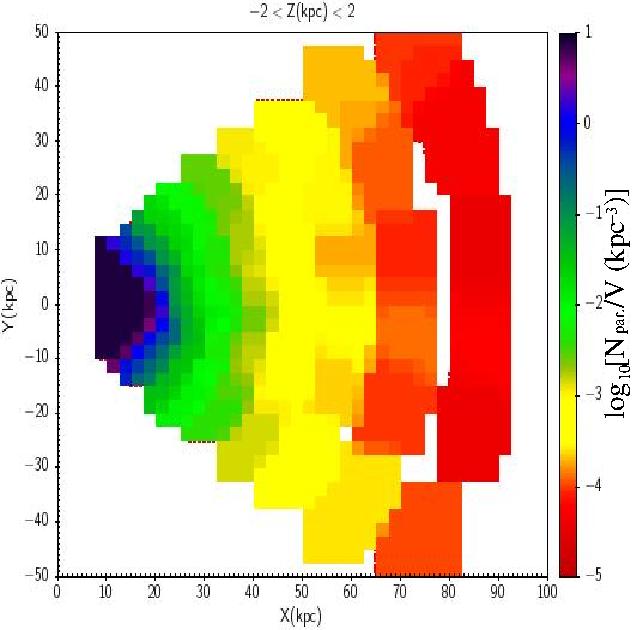}}
\par\centering
}
\caption{M-giant counts ($N{\rm param.}$) per unit volume. Here, we show the density without
correction of the selection effect or deconvolution. 
Pixel size 1x1 (kpc); interpolation in X and Y directions up to 5 pixels.}
\label{Fig:counts}
\end{figure}

\begin{figure}[htb]
\vspace{1cm}
{
\par\centering \resizebox*{8.5cm}{8.5cm}{\includegraphics{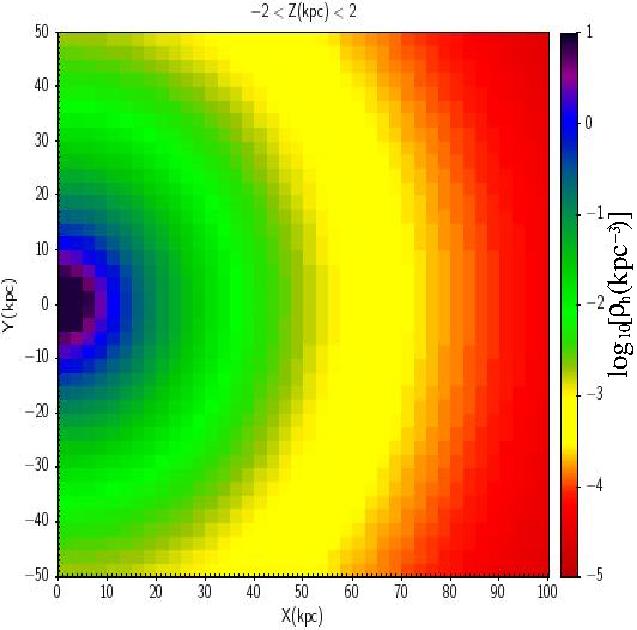}}
\par\centering
}
\caption{Prediction of [M/H]$>-1.5$ M-giants halo density from \citet{Fen89}, with $Q=10.093$, $\rho _\odot =2.5\times 
10^3$ star/kpc$^3$, and assuming that [M/H]$>-1.5$ stars are half of the total.}
\label{Fig:halomodel}
\end{figure}

\begin{figure}[htb]
\vspace{1cm}
{
\par\centering \resizebox*{8.5cm}{8.5cm}{\includegraphics{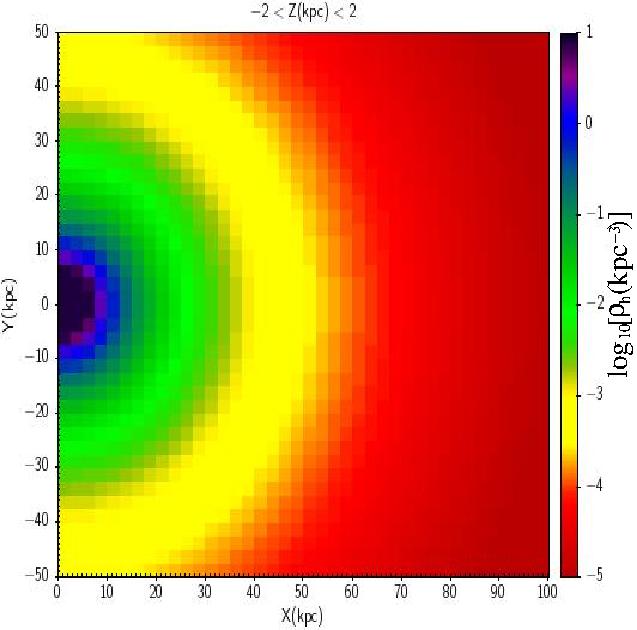}}
\par\centering
}
\caption{Prediction of [M/H]$>-1.5$ halo M-giants density from \citet{Xu18}, with $n=5.03$, $q=1$, $A_{X,\odot }=1.4\times 10^{-3}f\,\rho_\odot $, $\rho _\odot =2.5\times 
10^3$ star/kpc$^3$, and assuming that [M/H]$>-1.5$ stars are half of the total ($f=0.5$).}
\label{Fig:halomodel2}
\end{figure}

\begin{figure}[htb]
\vspace{1cm}
{
\par\centering \resizebox*{8.5cm}{8.5cm}{\includegraphics{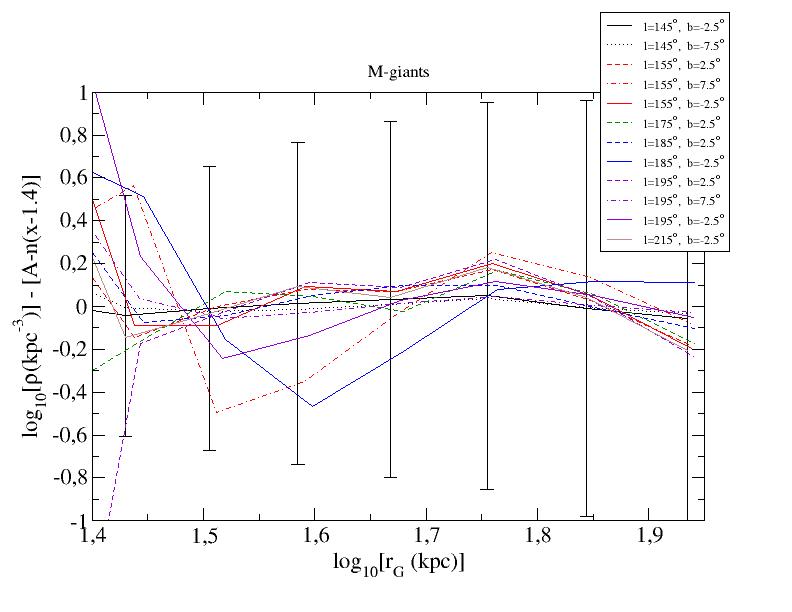}}
\vspace{1cm}
\par\centering \resizebox*{8.5cm}{8.5cm}{\includegraphics{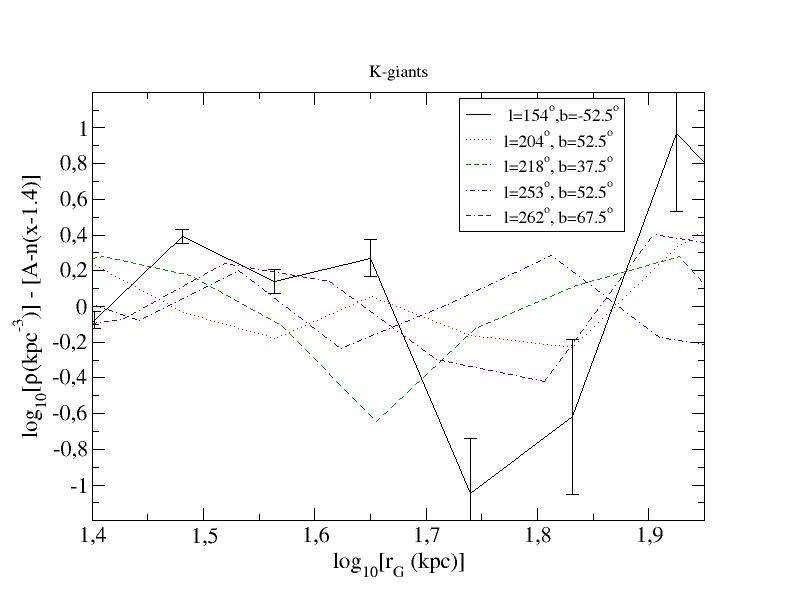}}
\par\centering
}
\caption{Residuals of the density given in Fig. \ref{Fig:density2} with respect to the linear fits for each line of 
sight (Table \ref{Tab:n}). Error bars (corresponding to the error of measured $\rho_0$) of the first line of sight are plotted; 
for other lines of sight, the error bars are similar.}
\label{Fig:residuals}
\end{figure}

\begin{figure}[htb]
\vspace{1cm}
{
\par\centering \resizebox*{8.5cm}{8.5cm}{\includegraphics{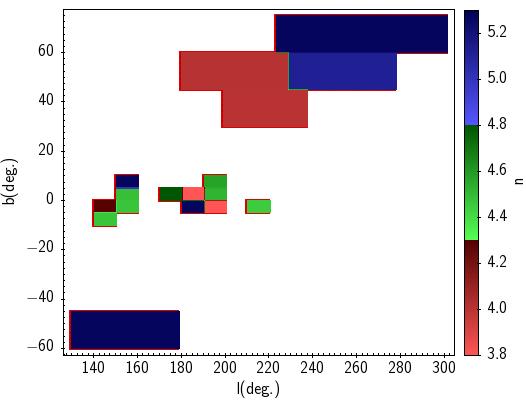}}
\par\centering
}
\caption{Variation of the power-law index $n$ with the sky position in the 17 investigated lines of sight.}
\label{Fig:slope_sky}
\end{figure}

\begin{figure}[htb]
\vspace{1cm}
{
\par\centering \resizebox*{8.5cm}{8.5cm}{\includegraphics{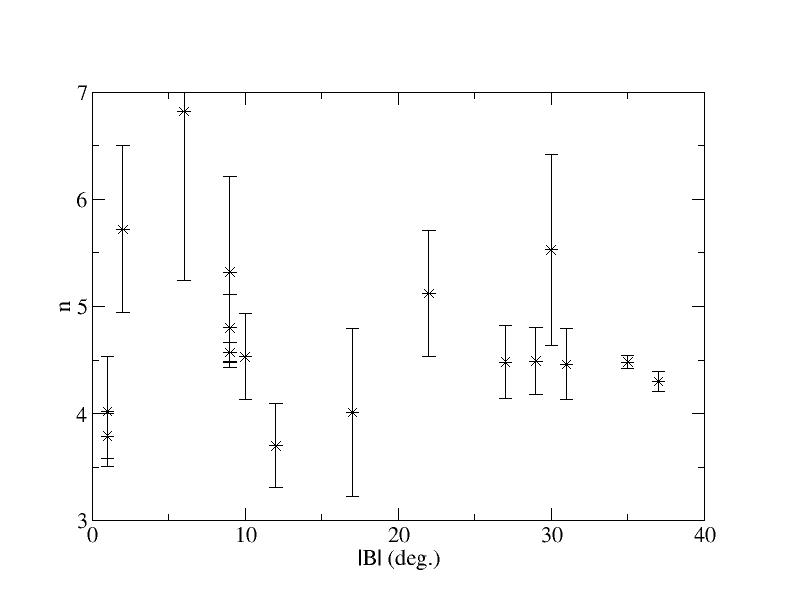}}
\par\centering
}
\caption{Power law index $n$ vs. angular distance to the Sagittarius stream orbit ($|\widetilde{B}|$).}
\label{Fig:Sagittarius}
\end{figure}

\end{document}